\documentclass[twocolumn,showpacs,preprintnumbers,amsmath,amssymb,superscriptaddress]{revtex4}

\usepackage{graphicx}
\usepackage{dcolumn}
\usepackage{bm}
\def\lsi{\raise0.3ex\hbox{$<$\kern-0.75em\raise-1.1ex\hbox{$\sim$}}}
\def\gsi{\raise0.3ex\hbox{$>$\kern-0.75em\raise-1.1ex\hbox{$\sim$}}}
\newcommand{\lsim}{\mathop{\lsi}}

\newcommand{\src}{\mbox{\footnotesize{src}}}
\newcommand{\obs}{\mbox{\footnotesize{obs}}}

\begin{document}
\title{An astronomical search for evidence of new physics: Limits on
    gravity-induced birefringence from the magnetic white dwarf RE
    J0317-853}
\author{Oliver Preuss}
\affiliation{Max-Planck-Institut f\"ur Aeronomie,
             D-37191 Katlenburg-Lindau, Germany}
\author{Mark P. Haugan}
\affiliation{Department of Physics,
             Purdue University 1396, West Lafayette, Indiana 47907, USA}
\author{Sami K. Solanki}
\affiliation{Max-Planck-Institut f\"ur Aeronomie,
         D-37191 Katlenburg-Lindau, Germany}
\author{Stefan Jordan}
\affiliation{Universit\"at T\"ubingen, Institut f\"ur Astronomie und Astrophysik, 
             D-72076 T\"ubingen, Germany}
\affiliation{Astronomisches Rechen-Institut, D-69120 Heidelberg, Germany}    
\date{\today}

\begin{abstract}
  The coupling of the electromagnetic field directly with gravitational 
  gauge fields leads to new physical effects that can be tested using 
  astronomical data. Here we consider a particular case for closer 
  scrutiny, a specific nonminimal coupling of torsion to electromagnetism,
  which enters into a metric-affine geometry of space-time. 
  We show that under the assumption of this nonminimal coupling, spacetime 
  is birefringent in the presence of such a gravitational field. 
  This leads to the depolarization of light emitted from extended 
  astrophysical sources. We use polarimetric data of the magnetic white 
  dwarf $\mbox{RE J0317-853}$ to set for the very first time constraints on 
  the essential coupling constant for this effect, giving $k^2 \lsim\, 
  (19 \,\mbox{m})^2 $.  
\end{abstract}
  
\maketitle
\section{Introduction}
  The designation \textit{new physics} is used here to refer to phenomena 
  beyond the scope of the standard model of particle physics or of general 
  relativity.  
  Today, efforts to develop a quantum theory of gravity or a complete, 
  consistent and unified theory of matter and all its interactions are 
  rich and compelling sources of speculation about new physics.  Certainly, 
  the effective field theories that emerge as low-energy limits of string 
  theories are littered with new fields and interactions.  Since cherished 
  symmetries like CPT and Lorentz invariance can be broken in 
  these contexts \cite{ck98,fn91}, high-precision experimental and 
  observational tests of these symmetries offer particularly effective ways 
  of searching for evidence of new physics.  

  In this letter we focus on consequences of interactions that couple the 
  electromagnetic field to new physics. In recent papers by Kostelecky \& 
  Mewes \cite{km01}, couplings of the electromagnetic field 
  to an axial-vector, $(k_{AF})^{\mu}$, and to a tensor, $(k_{F})^{\mu \nu 
  \sigma \tau}$, are used to provide a generic characterization of 
  conceivable new physics influencing the photon sector.  The $k_{AF}$ and 
  $k_{F}$ structures could, for example, be constructs of vector and tensor 
  fields in an underlying theory.  If such fields acquire nonzero vacuum 
  expectation values, their couplings to the electromagnetic field 
  spontaneously break global Lorentz invariance. The $k_{AF}$ structure 
  breaks CPT as well.  
  While motivated by current speculations regarding fundamental physics, 
  the conceptual framework and formalism of the work by Kostelecky 
  and Mewes is that of Lagrangian-based local field theory.  With different 
  motivations, questions regarding the most general way in which new 
  fields may couple to the electromagnetic field have been addressed in 
  this framework before.  It is reassuring to find all of these studies 
  reaching equivalent conclusions.  

  Today's strong empirical foundation for general relativity 
  is a consequence of continuing efforts to probe for evidence of this sort 
  of new or alternative physics, see Will \cite{will93} for a review.    
  In this research setting the Lagrangian-based local field theory framework 
  is often called the Dicke framework \cite{will93}.  
  Theoretical alternatives to general relativity can feature gravitational 
  fields in addition to the symmetric second-rank tensor potential of 
  general relativity.  One distinguishes two classes of alternative theories 
  based on the way in which these gravitational fields couple to matter.  
  A theory is metric if it admits a representation in which matter fields 
  are all minimally coupled in the familiar way to a symmetric second-rank 
  tensor gravitational field alone, implying that auxiliary fields couple 
  only to that metric tensor field and to each other.  Otherwise, a theory 
  is nonmetric.  
  It is in the study of nonmetric theories which couple gravitational 
  fields directly to the electromagnetic field that one finds analogues of 
  the recent work by Kostelecky and Mewes.  The equivalent of their Lagrangian 
  density governing the dynamics of the electromagnetic field can be found in 
  Ni's $\chi g$ formalism \cite{ni77}. There, this Lagrangian density 
  is expressed in terms of a fourth-rank tensor field, $\chi^{\alpha \beta 
  \gamma \delta}$, constructed from a symmetric second-rank tensor field, 
  $g^{\mu \nu}$, and auxiliary gravitational fields, 
  \begin{equation}
    {\cal L}_{EM} =  \chi^{\alpha \beta \gamma \delta} 
    F_{\alpha \beta} F_{\gamma \delta}.  
  \end{equation}
  Ni \cite{ni84} noted that theories encompassed by this formalism 
  can predict birefringence and used pulsar polarization observations to 
  constrain this possibility twenty years ago.  

  Recognition that gravitation theories touted as viable alternatives 
  to general relativity could predict such birefringence stimulated 
  new interest in such possibilities in the 1990s and sharp constraints 
  were imposed on birefringence predicted by Moffat's non-symmetric 
  gravitation theory (NGT) \cite{sh96,s99,gab91,gab91a,hk95}. 
  The analysis of birefringence by Kostelecky and Mewes 
  \cite{km01} and that performed by Haugan and Kauffmann  
  \cite{hk95} in the general context provided by the $\chi g$ formalism 
  run parallel to  one another.  Indeed, the 3+1 decomposition of 
  the $\chi$ tensor made by Haugan and Kauffmann yields SO(3) tensors 
  equivalent to the $\kappa$ matrices of Kostelecky and Mewes, 
  \begin{equation}
    (\kappa_{DE})^{ij} = 2 \xi^{ij}, \,\, 
    (\kappa_{HB})^{ij} = 2 \zeta^{ij} ,\,\, 
    (\kappa_{DB})^{ij} = 2 \gamma^{ij}.  
  \end{equation}
  Kostelecky and Mewes restrict their attention to cases in which these 
  quantities are spacetime constants.  Haugan and Kauffmann allow them to 
  vary, reflecting the structure of fields involved in new physics that 
  are generated by localized sources.  Their work establishes that evidence 
  of such birefringence can reveal new physics even when the fields involved 
  do not acquire nonzero vacuum expectation values and global Lorentz 
  invariance is not spontaneously broken.  
  String-inspired speculation regarding new physics provides a compelling 
  motivation to continue searching for evidence of birefringence of the 
  kind analyzed in preceding references.  

  As additional motivation we offer a new example, drawn from the 
  gravitation physics literature, of a theoretical alternative to general 
  relativity that naturally predicts such birefringence if we assume a special
  nonminimal coupling of torsion to electromagnetism. 
  Besides Moffat's NGT this provides a new concrete test case against which one
  can gauge the precision of birefringence searches. In this context it is 
  important to note that the new version of Moffat's NGT \cite{m95} also provides a 
  valid alternative to general relativity which has overcome the technical 
  difficulties found by Damour et al. \cite{d92} and Clayton et al. \cite{c98}.
  The foil for general relativity we present in section II, is drawn from 
  the class of metric-affine theories of gravity \cite{h95}.  
  We hope that this example will prompt further work on gravity-induced 
  birefringence predicted by metric-affine theories and other alternatives 
  to general relativity; see \cite{r03}.  
  In this paper we also report new observational constraints on such 
  gravity-induced birefringence based on data from  white dwarfs  -- which 
  constitute  the most abundant stage of final stellar evolution -- possessing
  a strong magnetic field.

\section{Gravity-induced birefringence in metric-affine theories}
  Members of the class of metric-affine theories of gravity feature torsion 
  and/or nonmetricity gravitational fields in addition to a symmetric 
  second-rank tensor gravitational potential \cite{h95}.  
  The minimal or geometrical coupling scheme for coupling gravity to matter 
  in metric-affine theories does couple torsion and nonmetricity to some 
  matter fields so metric-affine theories are, in fact, nonmetric theories.  
  However, the minimal coupling scheme does not couple torsion or nonmetricity 
  to the electromagnetic field \cite{p97}. Only 
  nonminimally coupled metric-affine theories can conceivably predict birefringence.  

  Here, we focus entirely on the effect of nonminimal coupling to torsion. We suggest, 
  in addition to the conventional Maxwell Lagrangian, the additional nonminimal piece \cite{p03}
  \begin{equation}
     L_{EM} = k^2 *(T_\alpha \wedge F)\,T^\alpha \wedge F \quad , 
  \end{equation}
  see \cite{ih03} for other possibilities,
  where $k$ is a coupling constant with the dimension of length, $*$ denotes the 
  Hodge dual, $T$ denotes the torsion and $F$ the electromagnetic field, which 
  is related in the usual way to its potential $A$. This addition is 
  consistent with gauge invariance and, so, with charge conservation. It 
  corresponds to a term 
  \begin{equation}
    \delta \! {\cal L}_{EM} = 
    \delta \! \chi^{\alpha \beta \gamma \delta} F_{\alpha \beta} 
    F_{\gamma \delta}, 
  \end{equation}
  in the notation of Haugan and Kauffmann \cite{hk95}. This 
  correspondence permits us to infer consequences of the nonminimal 
  torsion coupling above from the results of the general analysis of 
  reference \cite{hk95}.  
  Since our immediate interest is the interpretation of stellar data, we focus 
  on the case of static, spherically symmetric metric-affine fields to explore 
  the dominant effects of solar torsion on the propagation of light from the 
  Sun.  Tresguerres \cite{t95} has expressed the most general form of a static, 
  spherically symmetric torsion field in terms of six radial functions.  In 
  his notation these are $\alpha$, $\beta$ and $\gamma_{(i)}$ for $i=$1 to 4.  
  Plugging his representation into $\delta \! {\cal L}_{EM}$ above yields 
  \begin{eqnarray}
    \delta \! {\cal L}_{EM} &=& 
    k^2 \lbrace (\alpha^2 - \beta^2) B_1^2 
    -(\gamma_{(1)}^2 + \gamma_{(4)}^2) [B_2^2 + B_3^2]\nonumber \\ 
    &-&(\gamma_{(3)}^2 + \gamma_{(4)}^2) [E_2^2 + E_3^2] \\
    &+& 2(\gamma_{(1)} \gamma_{(4)} - \gamma_{(2)} \gamma_{(3)})[B_2 E_2 + B_3 E_3]\nonumber \\ 
    &+& 2(\gamma_{(1)} \gamma_{(3)} + \gamma_{(2)} \gamma_{(4)})[B_3 E_2 - B_2 E_3]\rbrace \nonumber
    \quad ,
  \end{eqnarray}
  where $E$ and $B$ refer in the usual way to the electric and magnetic 
  components of $F$ in Tresguerres' coordinates \cite{p03,sp03}. Note that the 
  three-vector index 1 refers to the radial direction. 
  In terms of the components of the $\xi$, $\zeta$ and $\gamma$ tensors of 
  reference \cite{hk95}, we can read off from the corresponding expression 
  for $\delta \! {\cal L}_{EM}$ how the nonzero components of the tensors 
  $\xi$, $\zeta$ and $\gamma$ are expressed in terms of the Tresguerres´ 
  functions for the nonminimal coupling considered here. To make a 
  specific prediction of the effect this 
  torsion coupling will have on the propagation of light from a star, we must 
  consider a specific metric-affine candidate for the stellar field.  We choose 
  a special case of a static, spherically symmetric field found by Tresguerres 
  \cite{t95a} in which nonmetricity vanishes, corresponding to vanishing 
  dilatation and shear charge. For simplicity we also set the cosmological constant 
  to zero. The corresponding forms of Tresguerres' $\alpha$, $\beta$ and 
  $\gamma_{(i)}$ functions are 
  \begin{eqnarray}
    \alpha (r) &=& {1 \over r} - {m \over r^2}\, ,\quad 
    \beta (r) =  {1 \over r} + {m \over r^2}\\
    \gamma_{(1)} (r) &=& - {1 \over 2} \left({1 \over r} + {m \over r^2}\right)\, , \,
    \gamma_{(3)} (r) =  {1 \over 2} \left({1 \over r} - {m \over r^2}\right) \, 
  \end{eqnarray}
  with $\gamma_{(2)} = \gamma_{(4)} = 0$.  A somewhat tedious 
  recapitulation of the analysis in \cite{hk95} in this case leads to 
  the conclusion that, to leading order, the specified nonminimal torsion 
  coupling and chosen stellar torsion field singles out linear polarizations 
  with a fractional difference in their propagation speeds of
  \begin{equation}\label{deltac}
    {{\delta c} \over c} = - \sqrt{6} \,{{k^2 m} \over r^3} \sin^2 (\theta ) ,
  \end{equation}
  where $\theta$ denotes the angle relative the outward radial direction in 
  which the light is propagating. In this context it is interesting to compare
  (\ref{deltac}) with equation (21) and (22) of \cite{r03}. 

  We are interested in the effect that this differential propagation has on 
  the polarization of light as it travels from a localized source on the 
  stellar surface to an observer.  This is determined by the phase shift 
  that accumulates between the polarization components singled out by the 
  stellar field as the light propagates. Recapitulating the analysis performed 
  in \cite{gab91,gab91a}, we find that in this case 
  \begin{equation}
    \Delta \Phi = \sqrt{2 \over 3} {{2 \pi k^2 m} \over {\lambda R^2}} 
    {{(\mu +2) (\mu - 1)} \over {\mu + 1}}, 
  \end{equation}
  where $\mu$ denotes the cosine of the angle between the line of sight
  and the normal on the stellar surface ($\mu=1$: stellar disk center, 
  $\mu=0$: limb), $\lambda$ is the light's wavelength and $R$ is the stellar 
  radius.  
  
\section{Data, analysis and results}
    All light received from a pointlike source of polarized radiation in a 
    gravitationally birefringent environment suffers the same phase 
    shift $\Delta \Phi(\mu)$. Introducing wavelength dependent Stokes 
    parameters $I_{\lambda},\,Q_{\lambda},\,U_{\lambda},V_{\lambda}$ to describe 
    polarized light \cite{s62}, with Stokes $Q$ defined to represent the 
    difference between linear polarization parallel and perpendicular 
    to the local stellar limb, gravitational birefringence introduces 
    a crosstalk between the linear polarization parameter Stokes $U$ 
    and the net circular polarization, $V$. 
    This crosstalk is such that although the observed values $U_{\obs}$ 
    and $V_{\obs}$ differ from the values emitted by the source, 
    $U_{\src}$ and $V_{\src}$, the composite degree of polarization 
    remains equal: $(U_{\obs}^2 + V_{\obs}^2)^{1/2} = (U_{\src}^2 + 
    V_{\src}^2)^{1/2}$. If an extended source covering a range of $\mu$ 
    values is observed then light emitted from different points suffers 
    different phase shifts and, so, adds up to an incoherent superposition. 
    Summing over the different contributions, while using the additive 
    properties of Stokes parameters, yields a reduction of the observed 
    polarization relative to the light emitted from the source:
    $(U_{\obs}^2 + V_{\obs}^2)^{1/2} < (U_{\src}^2 + V_{\src}^2)^{1/2}$. 
    Since the rotationally modulated polarization from magnetic white 
    dwarfs can only be produced by an extended source \cite{chile} any 
    observed (i.e. non-zero) degree of polarization provides a limit 
    on the strength of birefringence induced by the star's gravitational 
    field \cite{s99}. 

    The polarized radiation from white dwarfs is produced at the stellar 
    surface as a result of the presence of ultrastrong (up to $10^5$ T) 
    magnetic fields \cite{lan92}. 
    Since the disk of a white dwarf is unresolved, only the total 
    polarization from all surface elements is observable:
    \begin{equation}\label{vl}
      V_{\lambda, {\rm tot}}(k^2) = 2\pi\int\int \,V_{\lambda}(\mu,B,B_{\|})\,\cos(\Delta\Phi)
      \, \mu \, d\theta\, d\phi \,\, , 
    \end{equation}         
   
    \noindent where the Stokes parameter  $V_{\lambda}$ changes over the
    visible hemisphere and depends on the wavelength $\lambda$, the location $\mu$ 
    (limb darkening), the total magnetic field strength $B(\theta,\phi)$, the 
    line-of-sight component $B_{\|}(\theta,\phi)$, and on the parameters of the 
    stellar atmosphere. Gravitational birefringence reduces the polarization by
    means of $\cos(\Delta\Phi)$. Theoretically, the Stokes parameters can be 
    calculated by solving the radiative transfer equations through a magnetized 
    stellar atmosphere on a large number of surface elements on the visible hemisphere
    (e.g. \cite{j92}). If the star is rotating, the spectrum and polarization 
    pattern changes according to the respective magnetic field distribution 
    visible at a particular moment. To obtain the degree of circular polarization, 
    Eq.~(\ref{vl}) has to be divided by the total stellar flux $I_{\lambda, {\rm tot}}$ 
    emitted to the observer at wavelength $\lambda$. 
    Below we will calculate a maximum circular polarization $V_{\lambda, {\rm max}}/
    I_{\lambda, {\rm tot}}$ from radiative transfer calculations which is in general 
    higher than the observed value $V_{\lambda, {\rm obs}}/I_{\lambda, {\rm obs}}$. 
    Then we assume that the reduction from  $V_{\lambda, {\rm max}}$ to  $V_{\lambda, 
    {\rm obs}}$ is entirely due to the factor $\cos(\Delta\Phi(k^2))$ in Eq.~(\ref{vl}) 
    thereby calculating the maximum value for $k^2$, i.e. our limit on $k^2$ is reached 
    as soon as $V_{\lambda, {\rm tot}}/I_{\lambda, {\rm tot}}$ in Eq.~(\ref{vl}) becomes 
    smaller than $V_{\lambda,{\rm obs}}/I_{\lambda, {\rm tot}}$ for a certain value 
    of $k^2$. 
   
    $\mbox{RE J0317-853}$ is a remarkable object within the class of isolated 
    magnetic white dwarfs. Besides being the most rapidly rotating star 
    ($P=725$ sec) of this type it is also the most massive at $1.35\,M_{\odot}$, 
    close to the Chandrasekhar limit \cite{b95}. The corresponding radius 
    is only $0.0035\, R_{\odot}$. With a reported $V_{\lambda, {\rm obs}}/I_{\lambda, 
    {\rm tot}}$ of  $20$\%  at $\lambda= 576$ nm \cite{jb99}, $\mbox{RE J0317-853}$ 
    is also the magnetic 
    white dwarf with the highest known level of circular polarization. Due 
    to its small radius and high degree of circular polarization, $\mbox{RE J0317-853}$ is a 
    very suitable object for setting limits on gravitational birefringence.
    The analysis of time resolved UV flux spectra obtained with the Hubble Space 
    telescope has shown that the distribution of the field moduli
    is approximately  that of an off-centered magnetic dipole oriented obliquely to 
    the rotation axis with a polar field strength of $B_d=3.63\cdot10^4$ T, 
    leading to visible surface field strengths between $1.4\cdot 10^4$ T and 
    $7.3\cdot 10^4$ T \cite{bjo99}. This model is not only able to describe the UV, 
    but also  the optical spectra (Jordan et al. in prep.), which means that the 
    distribution of the magnetic field moduli - but not necessaryly of the longitudinal 
    components -  is correctly described. This result is completely independent of the 
    magnitude of the gravitational birefringence. 
    From radiative transfer calculations it follows that at the phase of rotation
    when the maximum value of 20\%\ polarization at 576 nm is measured, almost the 
    entire visible stellar surface is covered by magnetic fields between $1.4\cdot 10^4$
    and $2.0\cdot 10^4$ T, with only a small tail extending to maximum field 
    strengths of $5.3\cdot 10^4$. This distribution can best be modeled by 
    assuming a rotational phase where the axis of the off-centered dipole
    is nearly perpendicular to the line of sight. Using this special field geometry 
    we calculated a histogram distribution of the visible surface magnetic field 
    strengths in order to set sharp limits on gravitational birefringence.     
    For each field strength bin of the histogram we calculated the maximum circular
    polarization from radiative transfer calculations by assuming that the field vector 
    always points towards the observer. The total maximum polarization from the whole 
    visible stellar disk without gravitational birefringence is then calculated by 
    adding up the contributions from each field strength bin weighted with its relative 
    frequency. This results in $V_{\lambda, {\rm max}}/I_{\lambda,{\rm tot}} = 26.5$\%. 
    Assuming that the reduction to $V_{\lambda, {\rm obs}}/I_{\lambda,{\rm tot}}= 20$\% 
    is entirely due to gravity induced depolarization -- and not due to the fact that 
    in reality not all field vectors point towards the observer -- we find an upper 
    limit for this effect of $k^2 \lsim (19\,{\rm m})^2$. Since there is always a small
    uncertainty in determing the exact mass of a white dwarf, we also calculated an upper 
    limit on $k^2$ assuming a lower mass of $1\,M_{\odot}$. This leads to 
    $k^2 \lsim (22\,{\rm m})^2$.    
    An even more extreme assumption would be to use the maximum circular polarization 
    predicted by all field strengths in the interval $1.4 \cdot 10^4-5.3 \cdot 10^4$ (reached at
    $5.3\cdot 10^4$ T) for the whole stellar disk. Then we obtain $V_{\lambda, {\rm max}}/I_{\lambda, 
    {\rm tot}}=48.3$\%\ and an upper limit of $k^2 \lsim (30.5\,{\rm m})^2$.
    Independent from any dipole model and without any reference to radiative transfer calculations 
    the assumption of 100\%\ emerging polarization leads to $k^2 \lsim (45\,{\rm m})^2$.
    In order to compare the quality of our method with previous similar results from Solanki 
    et al. \cite{s99} we also set new upper limits on the NGT parameter $\ell^2_{\star}$ wich 
    also causes gravitational birefringence in case of $\ell^2_{\star}\neq 0$. Assuming $V_{\lambda, 
    {\rm max}}/I_{\lambda,{\rm tot}} = 26.5$\% leads to $\ell^2_{\star} \lsim (1.8\,{\rm km})^2$
    in contrast to $\ell^2_{\star} \lsim (4.9\,{\rm km})^2$, determined for the white dwarf
    GRW $+70^{\circ}8247$.    
\section{Conclusions}
  We have used spectropolarimetric observations of the massive 
  $(1.35\,M_{\odot})$ white dwarf $\mbox{RE J0317-853}$, to impose new strong 
  constraints on the birefringence of space-time in the presence of a 
  gravitational field. Such an effect is predicted by new physics in the 
  form of a direct coupling of the electromagnetic field to auxiliary 
  gravitational fields. Since the gravity-induced birefringence of 
  space violates the Einstein equivalence principle, our analysis also 
  provides a test of this foundation of general relativity and other 
  metric theories of gravity. We consider as a specific case a metric-affine 
  theory that couples the electromagnetic field with torsion and for 
  which a static spherically symmetric field has been found \cite{t95a}. 
  The data provide an upper limit for the relevant coupling constant $k^2$ of
  $(19\,\mbox{m})^2$ -- or $(45\,\mbox{m})^2$ for the most conservative 
  assumptions on. Considerably tighter limits based on the same 
  astronomical source could be provided by measurements of circular 
  polarization in the FUV (in particular associated with Ly$\alpha$ 
  absorption features) and also by a consistent model for the magnetic 
  field geometry which reproduces the spectropolarimetry measurements in 
  the optical.
  
 {\em We are grateful to Dayal T. Wickramasinghe and Friedrich W. Hehl
      for helpful remarks and valuable discussions.
    This research has made use of NASA's Astrophysics Data System 
      Abstract Service.}


\begin{thebibliography}{99}
  \bibitem{ck98}  D. Colladay, V.A. Kostelecky, Phys. Rev. D \textbf{58}, 116002 (1998) and 
                  references therein.  
  \bibitem{km01}  V.A. Kostelecky, M. Mewes, Phys. Rev. Lett. \textbf{87}, 251304 (2001)                     
                  and Phys. Rev. D \textbf{66}, 056005 (2002)             
  \bibitem{fn91}  C.D. Froggatt, H.B. Nielsen, Origin of Symmetries, World Scientific, Singapore, 1991
  \bibitem{will93}C.M. Will, \textit{Theory and Experiment in Gravitation Physics, Second 
                  Edition}, (Cambridge, 1993).  
  \bibitem{ni77}  W.-T. Ni, Phys. Rev. Lett. \textbf{38}, 301 (1977).  
  \bibitem{ni84}  W.-T. Ni, in \textit{Precision Measurements and Fundamental Constants II}, 
                  U.S. National Bureau of Standards 
                  Publication 617 (U.S. GPO, Washington D.C., 1984).  
  \bibitem{sh96}  S.K. Solanki, M.P. Haugan, Phys. Rev. D \textbf{53}, 997 (1996) 
  \bibitem{s99}   S.K. Solanki, M.P. Haugan, R.B. Mann, Phys. Rev. D \textbf{59}, 047101 (1999).
  \bibitem{gab91} M.D. Gabriel et al., Phys. Rev. D. \textbf{43}, 308 (1991).             
  \bibitem{gab91a} M.D. Gabriel {\em et al.}, Phys. Rev. D. \textbf{43}, 2465 (1991).
  \bibitem{hk95}  M.P.Haugan, T.F. Kauffmann, Phys. Rev. D \textbf{52}, 3168 (1995).
  \bibitem{m95}   J.W. Moffat, Phys. Letts. B \textbf{335}, 447 (1995)
  \bibitem{d92}   T. Damour, S. Deser, J. McCarthy, Phys. Rev. D \textbf{45}, R3289 (1992).
  \bibitem{c98}   M. Clayton, L. Demopoulos, J. Legare, GRG \textbf{30}, 1501 (1998).
  \bibitem{h95}   F.W. Hehl {\em et al.}, Phys. Repts. 258, 1 (1995).
  \bibitem{r03}   G.F. Rubilar, Y.N. Obukhov, F.W. Hehl, Class. Quantum Grav. {\bf 20} L185 
                  (2003), Preprint \texttt{gr-qc/0305049}.
  \bibitem{ih03}  Y. Itin, F.W. Hehl, Preprint \texttt{gr-qc/0307063}.            
  \bibitem{p97}   R.A. Puntigam, C. Laemmerzahl, F.W. Hehl, Class. Quantum Grav. \textbf{14}, 
                  1347 (1997). 
  \bibitem{t95}   R. Tresguerres, Z. Phys. C \textbf{65}, 347 (1995).
  \bibitem{t95a}  R. Tresguerres, Phys. Lett. A  \textbf{200}, 405 (1995).  
  \bibitem{s62}   W.A. Shurcliff, {\em Polarized Light}. Harvard University Press,
                  Harvard, Mass (1962).  
  \bibitem{chile} Magnetic Fields Across the Hertzsprung-Russell Diagram, ASP Conf. Proc. 
                  Vol. 248, p.479., 2001.
  \bibitem{lan92} J.D. Landstreet, Astron. Astrophys. Rev.\textbf{4}, 35 (1992) 
  \bibitem{aln73} C.W. Allen, {\em Astrophysical Quantities} (Athlone, 1973)  
  \bibitem{stift} Stift M.J., Mon. Not. R. Astron. Soc. 172, 133-139  (1975)   
  \bibitem{p03}   O. Preuss, PhD Thesis, Preprint \texttt{gr-qc/0305083}   
  \bibitem{sp03}  S.K. Solanki {\em et al.}, Phys. Rev. D {\bf 69} 062001 (2004)
  \bibitem{j92}   S. Jordan, Astron. Astrophys.\textbf{265}, 570 (1992)
  \bibitem{b95}   M.A. Barstow, S. Jordan, D. O'Donoghue, M.R. Burleigh, R. Napiwotzki, 
                  M.K. Harrop-Allin, Mon. Not. R. Astron. Soc. \textbf{277}, 971-985 (1995)
  \bibitem{bjo99} M.R. Burleigh, S. Jordan, W. Schweizer, Astrophys. J. \textbf{510}, L37-L40 (1999)
  \bibitem{jb99}  S. Jordan, M.R. Burleigh, Proc. of the 11th Europ. Workshop on White Dwarfs, 
                  ASP Conf. Ser. 169, p. 235 (1999)
  \end{thebibliography}
\end{document}